# Adaptive Plant Propagation Algorithm for Solving Economic Load Dispatch Problem

Sayan Nag[*]

Department of Electrical Engineering,
Jadavpur University
Kolkata, India

[*]Corresponding Author

*Abstract*—**Optimization problems in design engineering are complex by nature, often because of the involvement of critical objective functions accompanied by a number of rigid constraints associated with the products involved. One such problem is Economic Load Dispatch (ED) problem which focuses on the optimization of the fuel cost while satisfying some system constraints. Classical optimization algorithms are not sufficient and also inefficient for the ED problem involving highly nonlinear, and non-convex functions both in the objective and in the constraints. This led to the development of metaheuristic optimization approaches which can solve the ED problem almost efficiently. This paper presents a novel robust plant intelligence based Adaptive Plant Propagation Algorithm (APPA) which is used to solve the classical ED problem. The application of the proposed method to the 3-generator and 6-generator systems shows the efficiency and robustness of the proposed algorithm. A comparative study with another state-of-the-art algorithm (APSO) demonstrates the quality of the solution achieved by the proposed method along with the convergence characteristics of the proposed approach.**

I. INTRODUCTION

Optimization problems in Engineering Design, Structural Optimization, Economics and Scheduling Problems need mathematical models and Objective Function. These problems can be both Unconstrained (without constraints) and Constrained (with Constraints) and they involve continuous as well as discrete variables [1, 2, 3]. These problems as usually non-linear with a handful of constraints being active at the global optima making the task of finding the feasible solutions even more difficult. Optimization techniques and approaches have been applied to Stochastic Systems, Trajectory Systems, and Deterministic Systems. Stochastic Systems constitutes the optimal control problem in the presence of uncertainty. Trajectory Systems includes aerial paths but is primarily restricted to the computation of optimum rocket trajectories and space flight trajectories. Deterministic Systems involves the computations of the performance of mathematical models of various engineering designs for physical systems modelling with an appropriate objective function.

Economic Load Dispatch Problem (ED) is a problem of this kind involving both inequality and equality constraints with a pre-defined Objective Function. A variety of algorithms exist to solve this kind of problems – Exact and Approximate Algorithms. The Exact Algorithms include Quadratic Programming [4], Branch and Bound [5] and a few more whereas the Approximate Algorithms include Genetic Algorithms [6, 7, 8], Particle Swarm Optimization [9, 10], Colony Algorithms (Ant Colony [11, 12, 13, 14], Bee Colony [15, 16, 17]), Simulated Annealing [18, 19, 20], Tabu Search [21, 22, 23] and a lot more [24, 25, 26, 27]. These Colony Optimization Algorithms are also known as Memetic Algorithms. In general, these Approximate Algorithms are also known as Meta-heuristic Algorithms. Several methods including combination of one or two algorithms or combination of local search criterion or pool search techniques with these Approximate Algorithms are used throughout these years to solve multiple Constrained Optimization Problems.

___________________________________

[*]nagsayan112358@gmail.com

Economic Load Dispatch Problem (ED) is an important problem of research in the field of Power System because of the involvement of several non-linearity constraints. Keeping the total system generation cost minimum while satisfying all the existing equality and inequality constraints, the generation of each plant (unit) is to be determined within the limits (boundary constraints for each unit) to meet the requirements of the load demand. Essentially, the generator's active and reactive power are allowed to vary within a certain limit to meet a particular load demand so that the objective of keeping the fuel cost minimum is accomplished. This can be thought of to be similar to that of a Scheduling Problem like the Nurse Scheduling Problem. The Solution to the ED Problem devises a procedure for the distribution of total thermal generation requirements among several units for optimal system economy with incorporation of generating costs, transmission losses, and various acknowledged constraints imposed by the requisites of dependable service and equipment restraints.

The factors which influence the power generation at minimum cost are namely, the fuel cost transmission losses and generator efficiencies. It may happen that the generator which is the most efficient of all does not guarantee minimum cost because its location may be in an area where the cost of the fuel is high and also transmission losses tend to be much higher if the plant is at a location far away from the load center making the system uneconomical. The Classic Economic Load Dispatch Problem focusses on determination of the power output of every generating unit under the constraint condition of the system load demand keeping the net fuel generation cost minimum and neglecting line security constraints. The inequality constraints include: Voltage Constraints and Generator Constraints. In voltage constraints we have the upper and lower limits of the voltage magnitude and that of the load angle or the power angle. In generator constraints the KVA rating of the generator should not exceed the prescribed limits, thus there exists upper and lower boundaries for both the active power and the reactive power. The running spare capacity constraints are needed to meet forced outage of one or more generators and also to meet any unexpected load on the system. Transmission line constraints restrict the flow of power through the transmission lines to be less than its thermal capacity. There are also restrictions on the transformer tap set which depends on the transformer. We classify Economic dispatch area under one of the following four categories:

a) Optimal Power Flow
b) Dynamic Dispatch
c) Economic Dispatch in relation to Automatic Generation Control
d) Economic Dispatch with non-conventional generation sources

It is to be ensured that all load must be met in full as when they occur and with very high dependability to develop the distribution and expansion strategies of electrical utilities. The available facilities to store energy are few in number. So, the net production of a utility must closely track its load.

The Vector Optimization, commonly known as the Multi-objective Optimization gained importance because in the real world applications we encounter such problems where we have to deal with a number of objectives and they should be satisfied simultaneously for optimum operation. Economic Dispatch can be modelled as one such Multi-objective Optimization. Traditional methods including Gradient Descent, Dynamic Programming and Newton Methods are used earlier to solve the classical Economic Dispatch problem. But these traditional approaches are computationally inefficient. Also, keeping in mind the complexity of the non-linear optimization problems, the meta-heuristic algorithms such as GA, PSO, ABC, and ACO came into existence as mentioned previously. These meta-heuristic approximate algorithms provide better solutions to these optimization problems. Such an approximate optimization algorithm is Plant Propagation Algorithm (PPA) also known as Plant Propagation Algorithm [28, 29,30].

Plant Propagation Algorithm is a nature inspired new meta-heuristic algorithm which mimics the propagation of plants, particularly, the Strawberry Plant. Here we used a modified form of PPA known as the Adaptive-Plant Propagation Algorithm (APPA). The ED Problem is described in the next Section 2 of problem formulation. Section 3 contains the description of the APPA Algorithm. Section 4 describes the Test System. Section 5 contains the results and comparisons with other state-of-the-art approaches. Section 6 and Section 7 respectively contains the Conclusion and Future Scope and the References.

II. PROBLEM FORMULATION

Economic load dispatch (ED) can be defined as a non-linear problem where the goal is to minimize the total generation fuel cost while satisfying the load requirements and the constraints associated with the system simultaneously.

The net fuel cost function is considered as the important part of the cost of the overall power plant unit. This net cost comprises of three fundamental costs. It consists of the fixed costs such as the maintenance, wages, salaries of labors and interest and depreciation. They are therefore represented in the cost function as constant values. Next, the cost function is depended on the fuel cost, which is the most important part of the cost function, represented as a controllable objective in the total cost. Fuel cost is directly related to the active power output. In addition to this fuel cost, there exists a running cost which also has a direct relation with the square of the active power output. The net cost function for a particular unit is represented as a quadratic function as shown below:

$$Ci = Ci_1 + Ci_2 + Ci_3 \qquad (1)$$

where $C_i$ represents the total cost for the $i^{th}$ unit and $Ci_1$, $Ci_2$ and $Ci_3$ are respectively the fixed cost, fuel cost and running costs.

Now, considering all the generating units together, essentially this cost function can be represented as:

$$F = \sum_{i=1}^{N_g}(a_i + b_i P g_i + c_i P g_i^2) \qquad (2)$$

where $F$ represents the total cost including all the generating units, $a_i$, $b_i$ and $c_i$ are the coefficients or constants and $Pg_i$ represents the active power of the $i^{th}$ generating unit. Taking the sum of all the individual costs we get the overall cost where $N_g$ is the number of such generating units subjected to the constraints:

$$P_D = \sum_{i=1}^{N_g} Pg_i \qquad (3)$$

$$Pg_i^{min} < Pg_i < Pg_i^{max} \qquad (4)$$

A typical approach to solve the problem is to use the Lagrange multipliers to incorporate the constraints into the objective function. But this method is cumbersome and becomes more and more complex with the increase in number of variables or the generating units. Thus, meta-heuristics optimization approaches are used for solving the ED problem [34, 35, 36, 37, 38, 39, 40]. In our problem we considered the simplest case neglecting the losses with the assumption that the system is only one bus with all the generation and the loads connected to it, to show how our approach of Adaptive Plant Propagation Algorithm works for the Economic Load Dispatch (ED) problem.

III. PLANT PROPAGATION ALGORITHM (STRAWBERRY ALGORITHM)

Scientific studies in the recent years show that the plants also exhibit intelligent and fascinating behaviors [28, 30]. Plants do interact with the external world with inherent defense mechanisms to counteract the exploitation by insects like caterpillars. Root grows deep under the ground, the light and nutrient information goes to the growth center in root tips and the root is oriented accordingly. Hence, plants are really intelligent and thus attracts researchers and inspire them to develop algorithms based on their intelligence. Various algorithms exist like Flower Pollination Algorithm (FPA) [31, 32, 33], Root Shoot Growth Coordination Optimization (RSCO) and so on.

Unlike Simulated Annealing (SA), the Strawberry Algorithm or the Plant Propagation Algorithm (PPA) is a multipath following algorithm [28, 29, 30]. Essentially, this means that PPA is not restricted to a single path, rather there are multiple paths in which the solution may be directed. Thus, it has better exploitation and exploration properties unlike Simulated Annealing (SA). Exploitation is the property by which an algorithm makes a search nearer to optimum solutions and exploration means the covering of the search region. The better these two properties an algorithm have, the better are the chances of getting a good solution.

Plants are intelligent species which propagate through runners. Plants optimize its survival based on certain information like availability of water, nutrients, light and toxic substances. For the time being we will specifically consider Strawberry plant for our purpose since our algorithm also has another name, Strawberry Algorithm, although this algorithm applies for any plant. If a plant is present in a place where its roots are in a good location under the ground with availability of enough nutrients and water then it will hardly leave the position for its survival as long as the concentration of water and minerals in that spot remains more or less as per the requirement. So there will be a tendency to send short runners giving new strawberry plants, thus, occupying the neighborhood as much as it can. Now, for another example, if a plant is in a place devoid of the basic requirements like light, water and nutrients then it will try to find a better spot for its offspring by sending few long runners further at longer distances for exploiting the distant neighborhoods to find an optimal place for its survival. Also the number of long runners are a few since it becomes expensive for the plant to send long runners especially when it is in a poor location with no nutrients and water. Environmental factors decides a place, whether it is good or bad, whether it has plentiful nutrients and water concentrations or not. These factors in turn reflect upon the growth of plants and their sustainability. This underlying propagation strategy developed over time for ensuring survival in species receives attention and this is how Strawberry Algorithm or in general, Plant Propagation Algorithm is developed which imitates the propagation of plants. A set of parameters and functions is required: population size, a fitness function, the number of runners to be created for each solution and the distance or the range of allowable distance for each runner.

A plant is considered to be in a location $Y_i$ where the dimension of the search space is given as $n$. So, essentially $Y_i = \{y_{i,j}, \text{for all } j = 1,\ldots,n\}$. Let the population size be $N_p$ which determines the number of strawberry plants to be used initially. It is known that Strawberry plants which are in poor spots propagate by sending long runners which are few in number (the process is known as exploration) and the plants which are in location with abundance of essential nutrients, minerals and water propagate by sending many short runners (the process is known as exploitation). Maximum number of generations considered is $g_{max}$ (stopping criterion in the algorithm) and maximum number of permissible runners per plant is $n_{max}$.

The objective function values at different positions $Y_i$, $i = 1,\ldots,N_p$ are calculated. These possible candidate solutions will be sorted according to their fitness values or fitness scores. Here the fitness is a function of value of the objective function considered. It is better to keep the fitness scores within a certain range between 0 and 1, that is, if the fitness function is $f(x) \in [0, 1]$. To keep the fitness values within this range a mapping is done which involves another function described as:

$$N(x) = \frac{1}{2}(\tanh(4 f(x) - 2) + 1) \tag{5}$$

The number of runners that are determined by the functions and the distance of propagation of each of them are described. There exists a direct relation between the number of runners produced by a candidate solution and its fitness given as:

$$n_r = ceil(n_{max} \, N_i \, r) \tag{6}$$

Here, $n_r$ is the number of runners produced for solution $i$ in a particular generation or iteration after the population is sorted according to the fitness given in (5), $n_{max}$ is the number of runners which is maximum permissible, $N_i$ is the fitness as shown in (5) and $r$ is just a random number lying between 0 and 1 which is randomly selected for each individual in every iteration or generation. The minimum number of runners is 1 and maximum is $n_r$. This ensures that at least 1 runner should be there which may correspond to the long runner as described before. The length of each runner is inversely related to its growth as shown below:

$$d_j^i = (1 - N_i)(r - 0.5), \text{ for } j = 1,\ldots, n \tag{7}$$

where, $n$ represents the dimension of the search space. So, each runner is restricted to a certain range between -0.5 and 0.5. The calculated length of the runners are used to update the solution for further exploration and exploitation of the search space by the equation:

$$x_{i,j} = y_{i,j} + (b_j - a_j) d_j^i, \text{ for } j = 1,\ldots, n \tag{8}$$

The algorithm is modified to be an adaptive one based on the boundaries of the search domain. Hence, the name is given as Adaptive Plant Propagation Algorithm (APPA) or Strawberry Algorithm (ASA). If the bounds are violated the point is adjusted to lie within the search space. Essentially, $a_j$ and $b_j$ are the respective lower and upper bounds of the $j^{th}$ coordinate of the search space. After all individual plants in the population have conveyed their allotted runners, new plants are assessed and the entire expanded population is sorted. To keep the population fixed, rather the size of the population fixed, it is to be ensured that the candidates with lower growth are eliminated from the population. Another strategy is adopted to avoid being struck in the local minima. It may happen that for a certain number of consecutive generations there is no improvement in a candidate solution, rather the runners it sends out are also not fit to remain in the population. So a threshold to be set for such a solution such that if the number of generations in which it is not improving exceeds the threshold then the solution is abandoned and a new fresh candidate solution or individual (can be seen as a new plant) is generated within the bounds of the search space.

In the Economic Load Dispatch (ED) problem, this proposed Adaptive Plant Propagation Algorithm (APPA) is applied with constraints as dictated by the equations (3) and (4) and cost function as described in equation (2). All these equality and inequality constraints are taken care of; the inequality constraints basically defines the search space of the approximate solution and the equality constraint is taken care of using the penalty method. The following picture is a representation of the algorithm with all the steps to be performed in detail.

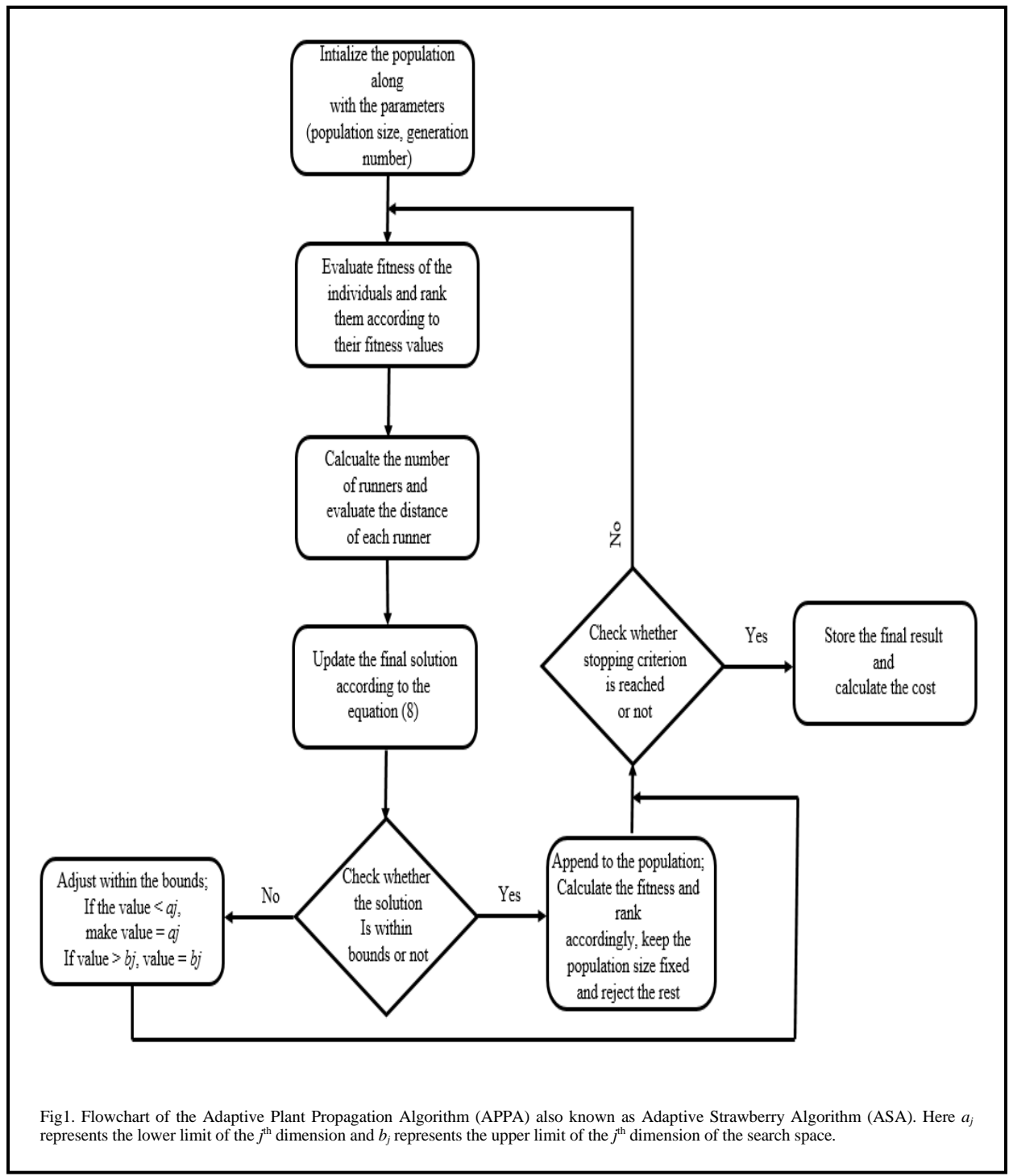

Fig1. Flowchart of the Adaptive Plant Propagation Algorithm (APPA) also known as Adaptive Strawberry Algorithm (ASA). Here $a_j$ represents the lower limit of the $j^{th}$ dimension and $b_j$ represents the upper limit of the $j^{th}$ dimension of the search space.

## IV. TEST SYSTEM

We considered two problems, both of them are Economic Load Dispatch Problem. In both the cases we considered a simple system neglecting the transmission losses. Problem1 consists of 3 three thermal plants with power output of generators as $P_1$, $P_2$ and $P_3$ (all are in MWs). The total load demand is $P_D$ which is 800 MW. The limits of the generators are given in the following table. We showed the solutions that came from two methods: (i) analytical method using Lagrange Multipliers and (ii) our proposed method (APPA).

TABLE I.

| Serial Number of Generators ($i$) | Thermal Unit Coefficients of cost characteristic equation | | |
|---|---|---|---|
| | $A_i$ (Rs/MW²h) | $B_i$ (Rs/MWh) | $C_i$ (Rs/h) |
| 1. | 0.004 | 5.3 | 500 |
| 2. | 0.006 | 5.5 | 400 |
| 3. | 0.009 | 5.8 | 200 |

TABLE II.

| Serial Number of Generators ($i$) | Power generation Limits | |
|---|---|---|
| | Lower Limit (MW) | Upper Limit (MW) |
| 1. | 350 | 450 |
| 2. | 200 | 300 |
| 3. | 100 | 200 |

Here, Table1 shows the values of thermal coefficients of cost characteristic equation in accordance with the generators number and Table2 shows the lower and upper limits of each of the generators.

Problem2 consists of 6 three thermal plants with power output of generators as $P_1$, $P_2$, $P_3$, $P_4$, $P_5$ and $P_6$ (all are in MWs). The total load demand is $P_D$ which is 410 MW. The limits of the generators are given in the following table. We showed the solutions that came from two methods: (i) Adaptive Particle Swarm Optimization (APSO) and (ii) our proposed method (APPA).

TABLE III.

| Serial Number of Generators ($i$) | Thermal Unit Coefficients of cost characteristic equation | | |
|---|---|---|---|
| | $A_i$ (Rs/MW²h) | $B_i$ (Rs/MWh) | $C_i$ (Rs/h) |
| 1. | 0.15247 | 38.53973 | 756.79886 |
| 2. | 0.10587 | 46.15916 | 451.32513 |
| 3. | 0.02803 | 40.39655 | 1049.99770 |
| 4. | 0.03546 | 38.30553 | 1243.53110 |
| 5. | 0.02111 | 36.32782 | 1658.56960 |
| 6. | 0.01799 | 38.27041 | 1356.65920 |

TABLE IV.

| Serial Number of Generators (*i*) | Power generation Limits | |
|---|---|---|
| | *Lower Limit* (MW) | *Upper Limit* (MW) |
| 1. | 10 | 30 |
| 2. | 20 | 40 |
| 3. | 40 | 55 |
| 4. | 40 | 55 |
| 5. | 120 | 140 |
| 6. | 120 | 140 |

Here, Table3 shows the values of thermal coefficients of cost characteristic equation in accordance with the generators number and Table4 shows the lower and upper limits of each of the generators.

## V. RESULTS

We have tabulated the results for both the problems. The results are shown below. The results obtained by the proposed approach are compared with the traditional method using Lagrange multipliers for problem 1 and for problem 2 the results obtained by the proposed APPA approach are compared with that obtained by APSO approach. All the constraints are taken care of in the proposed algorithm which is reflected in the results.

TABLE V.

| Algorithm / Method Used | Generator Power Output ($P_i$) | | | Cost (Rs/hr) |
|---|---|---|---|---|
| | $P_1$ (MW) | $P_2$ (MW) | $P_3$ (MW) | |
| Analytical Approach Using Lagrange Multipliers | 400 | 250 | 150 | 6,682.5 |
| Proposed Approach (APPA) | 400.0421 | 248.2936 | 151.5611 | 6,681.7 |

TABLE VI.

| Algorithm / Method Used | Generator Power Output ($P_i$) | | | | | | Cost (Rs/hr) |
|---|---|---|---|---|---|---|---|
| | $P_1$ (MW) | $P_2$ (MW) | $P_3$ (MW) | $P_4$ (MW) | $P_5$ (MW) | $P_6$ (MW) | |
| APSO | 13.9565 | 38.5234 | 50.7257 | 44.5845 | 138.452 | 125.623 | $5.9210 \times 10^7$ |
| Proposed Approach (APPA) | 26.8728 | 29.5504 | 50.2616 | 40.1164 | 131.099 | 133.059 | $5.8136 \times 10^7$ |

Table 4 and Table 5 shows the results of Problem 1 and Problem 2 respectively. In case of Problem 1, the number of generations considered is 5 and population strength is 30 and for Problem 2, the number of generations considered is 100 and the number of populations is 1000.

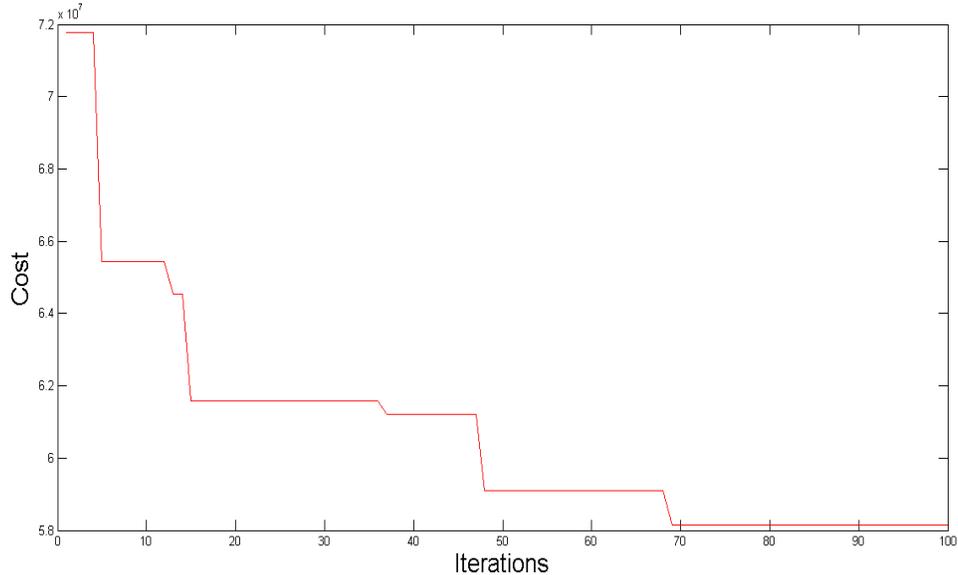
Figure 2 shows the convergence characteristics of total fuel cost and number of iterations for Problem 2. The proposed algorithm, APPA gives a robust solution and good convergence characteristics as shown in this figure.

Thus, it can be inferred that the proposed algorithm, APPA gives appreciable convergence characteristics along with a robust solution. Also, from the results it is clear that APPA and APSO more or less gives the same result, though cost-wise APPA is better than APSO in this case.

## VI. Conclusion And Future Scope

The optimal scheduling of an electric power system for determination of the generation for each unit such that the overall system generation cost is minimized while satisfying the system constrains is really an important problem in electric power sector. In this paper an investigation is done for the economic load dispatch problem solution using a novel technique called Adaptive Plant Propagation Algorithm (APPA) or Adaptive Strawberry Algorithm (ASA). The comparisons and results showed fascinating results regarding the efficiency and convergence characteristics of the proposed method thus making it fast for solving optimization problems. The future scope includes application of this approach to Multi-Objective Economic Load Dispatch Optimization Problems and consideration of the transmission losses. The proposed approach can be viewed upon as an efficient alternative method for solving the ED Optimization Problems. Modifications to the proposed methods may lead to efficient methods eventually producing fascinating solutions. Also, adjusting the penalty factors may lead to efficient solutions.

## VII. References

[1] K. Deb, "Optimization for Engineering Design: Algorithms and Examples", PHI Learning, 2004.

[2] X.-S. Yang, "Biology-derived algorithms in engineering optimization," in Handbook of Bioinspired Algorithms and Applications, S. Olarius and A. Y. Zomaya, Eds., chapter 32, pp. 589– 600, Chapman & Hall/CRC Press, 2005.

[3] L. C. Cagnina, S. C. Esquivel, and C. A. C. Coello, "Solving engineering optimization problems with the simple constrained particle swarm optimizer," Informatica, vol. 32, no. 3, pp. 319– 326, 2008.

[4] J. Z. Cha and R. W. Mayne, "Optimization with discrete variables via recursive quadratic programming—part 1: concepts and definitions," Journal of Mechanical Design, vol. 111, no. 1, pp. 124–129, 1989.

[5] E. Sandgren, "Nonlinear integer and discrete programming in mechnical design optimization," Journal of Mechanical Design, vol. 112, no. 2, pp. 223–229, 1990.

[6] S.-J. Wu and P.-T. Chow, "Genetic algorithms for nonlinear mixed discrete-integer optimization problems via meta-genetic parameter optimization," Engineering Optimization, vol. 24, no. 2, pp. 137–159, 1995.

[7] J. H. Holland and J. S. Reitman, "Cognitive systems based on adaptive algorithms," ACM SIGART Bulletin, no. 63, p. 49, 1977.

[8] K. Lamorski, C. Sławinski, F. Moreno, G. Barna, W. Skierucha, ´ and J. L. Arrue, "Modelling soil water retention using support vector machines with genetic algorithm optimisation," The Scientific World Journal, vol. 2014, Article ID 740521, 10 pages, 2014.

[9] R. Eberhart and J. Kennedy, "A new optimizer using particle swarm theory," in Proceedings of the 6th International Symposium on Micro Machine and Human Science (MHS '95), pp. 39– 43, IEEE, Nagoya, Japan, October 1995.

[10] G.-N. Yuan, L.-N. Zhang, L.-Q. Liu, and K. Wang, "Passengers' evacuation in ships based on neighborhood particle swarm optimization," Mathematical Problems in Engineering, vol. 2014, Article ID 939723, 10 pages, 2014.

[11] M. Dorigo and G. D. Caro, "Ant algorithms for discrete optimization", Artificial Life, vol. 5, no. 3, (1999), pp. 137-172.

[12] J. Dero and P. Siarry, "Continuous interacting ant colony algorithm based on dense hierarchy", Future Generation Computer Systems, vol. 20, no. 5, (2004), pp. 841-856.

[13] M. Dorigo and L. M. Gambardella, "Ant colony system: a cooperative learning approach to the traveling salesman problem", IEEE Transactions on Evolutionary Computation, vol. 1, no. 1, (1997), pp. 53-66.

[14] L. M. Gambardella, E. Taillard and M. Dorigo, "Ant colonies for the quadratic assignment problem", Journal of the Operational Research Society, vol. 50, no. 1, (1999), pp. 167-176.

[15] D. Karaboga and B. Akay. A comparative study of artificial bee colony algorithm. Applied Mathematics and Computation, 214(1):108–132, 2009.

[16] JC Bansal, H Sharma and SS Jadon "Artificial bee colony algorithm: a survey." Int. J. of Advanced Intelligence Paradigms 5.1 (2013): 123-159.

[17] Singhal, Prateek K., et al. "A new strategy based artificial bee colony algorithm for unit commitment problem." Recent Developments in Control, Automation and Power Engineering (RDCAPE), 2015 International Conference on. IEEE, 2015.

[18] C. Zhang and H.-P. Wang, "Mixed-discrete nonlinear optimization with simulated annealing," Engineering Optimization, vol. 21, no. 4, pp. 277–291, 1993.

[19] S. Anand, S. Saravanakumar and P. Subbaraj, "Customized Simulated Annealing based Decision Algorithms for Combinatorial Optimization in VLSI Floorplanninng Problem", Computational Optimization and Application, Vol. 30, No. 2, pp. 667-689, 2010.

[20] E. H. L. Aarts, J. H. M. Korst, and P. J. M. van Laarhoven, "Simulated annealing," in Local Search in Combinatorial Optimization, pp. 91–120, 1997.

[21] D. Clark(1993). "Exam scheduling by Tabu Search". Australian Soc. Ops Res. Bulletin, 12, 5–9.

[22] Glover F., Taillard E. andde Werra D. (1993). "A user's guide to Tabu Search". Annals of Ops Res., 41, 3–28.

[23] A. Hertz(1991). "Tabu search for large scale timetabling problems". Eur. J. Opl Res., 54, 39–47.

[24] Shah-Hosseini, "The intelligent water drops algorithm: a nature-inspired swarm-based optimization algorithm", International Journal of Bio-Inspired Computation 1(1/2), 71–79 (2009).

[25] K. Krishnanand, D. Ghose, "Glowworm swarm optimization for simultaneous capture of multiple local optima of multimodal functions", Swarm Intelligence 3(2), 87–124 (2009)

[26] X.S. Yang, "Firefly algorithms for multimodal optimization", LNCS, vol. 5792, pp. 169–178. Springer, Heidelberg (2009).

[27] M. Jaberipour and E. Khorram, "Two improved harmony search algorithms for solving engineering optimization problems," Communications in Nonlinear Science and Numerical Simulation, vol. 15, no. 11, pp. 3316–3331, 2010.

[28] X.-S. Yang, Nature-Inspired Metaheuristic Algorithms, Luniver Press, 2011.

[29] A. Salhi and E. S. Fraga, "Nature-inspired optimisation approaches and the new plant propagation algorithm," in Proceedings of the The International Conference on Numerical Analysis and Optimization (ICeMATH '11), Yogyakarta, Indonesia, 2011.

[30] J. Brownlee, Clever Algorithms: Nature-Inspired Programming Recipes, 2011.

[31] X-S Yang (2012), "Flower Pollination Algorithm for global optimization", Unconventional computation and natural computation. Springer. pp 240–249.

[32] XS Yang, M Karamanoglu, X He (2013) "Multi-objective flower algorithm for optimization", Procedia Comput Sci 18:861–868.

[33] XS Yang, M Karamanoglu, X He (2014), "Flower pollination algorithm: a novel approach for multiobjective optimization", Eng Optim 46:1222–1237.

[34] LDS Coelho, VC Mariani, "Particle swarm approach based on quantum mechanics and harmonic oscillator potential well for economic load dispatch with valve-point effects", Energy Convers Manage 2008;49(11):3080–5.

[35] Chaturvedi, Pandit, Srivastava. "Self-organizing hierarchical particle swarm optimization for nonconvex economic dispatch", IEEE Trans Power System 2008;23(3):1079–87.

[36] JB Park, KS Lee, JR Shin, KY Lee. "A particle swarm optimization for economic dispatch with nonsmooth cost functions", IEEE Trans Power System 2005;20(1):34–42.

[37] A Chatterjee, SP Ghoshal, V Mukherjee. "Solution of combined economic and emission dispatch problems of power systems by an opposition-based harmony search algorithm". Int J Electr Power Energy Syst 2012;39(1):9–20.

[38] GC Liao. "Integrated isolation niche and immune genetic algorithm for solving bid-based dynamic economic dispatch". Int J Electr Power Energy System 2012;42(1):264–75.

[39] C Peng, H Sun, J Guo, G Liu. "Dynamic economic dispatch for wind-thermal power system using a novel bi-population chaotic differential evolution algorithm". Int J Electr Power Energy System 2012;42(1):119–26.

[40] S Titus, AE Jeyakumar. "A hybrid EP-PSO-SQP algorithm for dynamic dispatch considering prohibited operating zones". Electr Power Compon System 2008;36(5):449–67.